\documentclass[twocolumn]{revtex4}
\usepackage{graphicx,amsmath}

\def\et{{\it et al.\/}}

\catcode`\é=13\def é{\'e} \catcode`\è=13\def è{\`e}

\catcode`\ù=13\def ù{\`u} \catcode`\à=13\def à{\`a}

\catcode`\ê=13\def ê{\^e} \catcode`\â=13\def â{\^a}

\catcode`\ô=13\def ô{\^o} \catcode`\û=13\def û{\^u}

\catcode`\î=13\def î{\^\i} \catcode`\ç=13\def ç{\c{c}}

\catcode`\ï=13\def ï{\"{i}}
\begin{document}

\title { Proximity DC squids in the long junction limit}
\author{L. Angers, F. Chiodi, J. C. Cuevas, G. Montambaux, M. Ferrier, S. Gu\'eron, H. Bouchiat}
\affiliation{Laboratoire de Physique des Solides, Associ\'e au CNRS, B\^atiment 510, Universit\'e Paris-Sud, 91405, Orsay, France}
\today

\begin{abstract}
We report the design and measurement of Superconducting/normal/superconducting (SNS) proximity DC squids in the long junction limit, i.e. superconducting loops interrupted by two normal metal wires roughly a micrometer long. Thanks to the clean interface between the metals, at low temperature a large supercurrent flows through the device. The dc squid-like geometry leads to an almost complete periodic modulation of the critical current through the device by a magnetic flux, with a flux periodicity of a flux quantum h/2e through the SNS loop.  In addition, we examine the entire field dependence, notably the low and high field dependence of the maximum switching current. In contrast with the well-known Fraunhoffer-type oscillations typical of short wide junctions, we find a monotonous gaussian extinction of the critical current at high field. As shown in \cite{Cuevas}, this monotonous dependence is typical of long and narrow diffusive junctions. We also find in some cases a puzzling reentrance at low field.  In contrast, the temperature dependence of the critical current is well described by the proximity effect theory, as found by Dubos {\it et al.} \cite{Dubos} on SNS wires in the long junction limit. The switching current distributions and hysteretic IV curves also suggest interesting dynamics of long SNS junctions with an important role played by the diffusion time across the junction.
\end{abstract}

\maketitle

\subsection{Introduction}

The proximity effect, or the penetration of superconducting correlations in a normal (non superconducting) conductor due to the proximity to a superconducting one, has been shown to give the normal conductor superconducting-like properties over a length that can be considerable at low temperatures. This is because at low temperature both the phase coherence length and the thermal length $\sqrt{\hbar D/k_BT}$ can extend beyond the length of the normal metal. This quantum coherence of a mesoscopic conductor then ensures that Andreev pairs, electron-hole time-reversed pairs formed at the SN interface, can propagate over distances of the order of the phase coherence length in the normal metal. Such a length is notably greater than the superconducting coherence length $\sqrt{\hbar D/\Delta}$. The normal conductors have ranged from simple noble metals to semi-conducting planes or wires, to long or short molecules (carbon nanotubes , endohedral fullerenes, DNA ) down to individual atoms in break junction geometries \cite{Kasumov,Scheer,VB,Silvano}.

One of the most remarkable consequence of the proximity effect is the propensity of a normal conductor well connected to two superconductors with different phases to carry a supercurrent, thereby demonstrating that Andreev pairs carry superconducting correlations along with information about the macroscopic phase of the superconductor at the boundary. Depending on the length of the normal part, the maximal supercurrent (called critical current $I_c$) that can be carried is determined either by the superconducting gap of the superconductor, or by the characteristic energy of the normal metal, the so called Thouless Energy, which in a diffusive conductor of diffusion constant $D$ is given by $E_{Th}=\hbar D/{L^2}$. Indeed the critical current has been shown \cite{Kulik} to be given by $eR_NI_c(T=0) = 2.07 \Delta$ in the case of short junctions ($L<\xi_S$, or equivalently $E_{Th}>\Delta$), and $eR_NI_c(T=0) = 10.82 E_{Th}$ in the opposite limit of long junctions. More generally, the remarkable point is that in the short junction limit the superconductor determine the properties of the proximity system, and in the long junction limit it is the normal metal (via its length, diffusion time, coherence length) which determines the proximity effect.

The interplay of the mesoscopic phase coherence in the normal metal and the superconducting coherence due to the nature of the superconducting state, have led to the fabrication of many interferometer devices. 
The experiments that launched the development of the mesoscopic proximity effect mainly consisted of a normal metal loop, with nearby superconducting islands (\cite{Courtois,Petrashov}). The resistance of these devices was modulated by the magnetic field one or two orders of magnitude more than the universal conductance fluctuations in normal rings with no superconductor. But no supercurrent was measured because normal leads connected the mesoscopic devices.
Interferometers in which a supercurrent was obtained and modulated are much more recent, and in most cases do not use a noble metal as the normal part: the non superconducting part actually is made of a long molecule such as carbon nanotubes\cite{VB}, or semiconducting wires\cite{Silvano}. In some cases $\pi$ junctions can be produced, either by the use of a ferromagnet between superconductors or by the action of a gate electrode which modulates the population of the normal part.

Interestingly, although many mesoscopic interferometers have been devised with a normal metal between several superconductors (\cite{Petrashov, Courtois}), no equivalent to a dc-squid (a superconducting loop with two insulating barriers or weak links) has yet been fabricated in the mesoscopic regime \cite{Clarke}. In this article, we describe several proximity dc-squids, i.e. superconducting loops with two normal metal bridges,  ranging from 0.75 to 1.9 microns long. We find that the IV curves are hysteretic, a phenomenon that we attempt to explain by the intrinsic dynamics of the diffusive normal metal. We find that the temperature dependence of the critical current of each long junction is in accordance to the theory of the proximity effect,  much like in the work of Dubos et al. \cite{Dubos}. The field dependence is more original: In the proximity dcsquid configuration the critical current is highly modulated by the magnetic field. In both the single SNS junction and the dcsquid configuration, at large perpendicular magnetic field, the extinction of critical current with field is monotonous and does not show the interference pattern previously observed in wider junctions \cite{Heida,Kutchinsky,graphene}. We present two theoretical models which attempt to explain this behaviour by including the diffusive nature of the normal metal. A surprising reentrance of the critical current at low field is also observed in some samples, and remains unexplained.

\subsection{Sample fabrication and measurement apparatus}
\begin{figure} 
\begin{center} 
\includegraphics[clip=true,width=7cm]{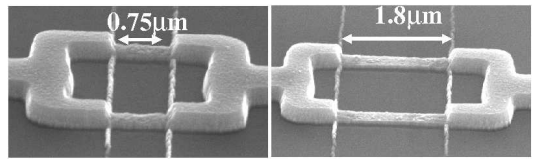}
\end{center}
\caption{SEM pictures of the two Nb/Au/Nb SNSdcsquids.}
\label{Photos} 
\end{figure}

\begin{figure} 
\begin{center} 
\includegraphics[clip=true,width=7cm]{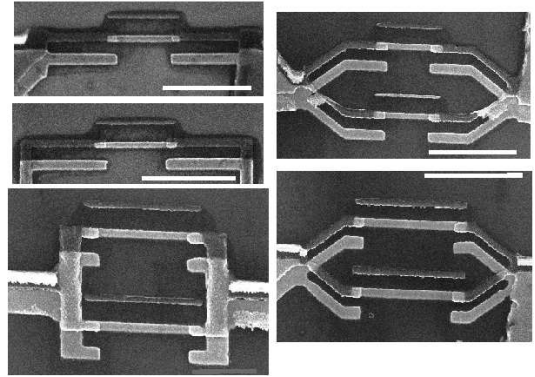}
\end{center}
\caption{SEM pictures of Al/Au/Al SNS wires and SNS dc squids: from top to bottom, and left to right: wires Alw1250 and Alw1300, SNS dc squids Alsq1500, Alsq900 and Alsq1900. Scale bar is 2 $\mu m$.}
\label{SEMAl} 
\end{figure}

As a normal metal we used $99.9999 \% $ pure gold, with a content in magnetic impurities (Fe) smaller than $0.1$ part per million.
Weak localization measurements in a simple wire made of this same material (at a different time) determined a phase coherence length  of the order of $10~\mu$m  below 50 mK \cite{Billangeon}. Two different superconductors were used, either aluminum or niobium (see fig. \ref{Photos} and \ref{SEMAl}).
 The proximity wires and dc squids made from Al and Au were obtained by conventional double angle thermal evaporation through a suspended mask, in a vacuum of $10^{-6}$ mbar. The process with niobium as a superconductor is more involved. We start with a bilayer of 50 nm-thick thermally evaporated gold, covered by a 200 nm-thick layer of sputtered niobium, and cover the bilayer with a 70 nm-thick aluminum layer. With e-beam lithography we define and wet-etch open a window in the Al mask through which we locally etch away the Nb using an SF6 plasma. In a realignment step we create a ring or wire shaped Al mask around which the bilayer is removed by an SF6 reactive ion etch followed by an argon ion beam etch. Finally the Al mask is removed and the samples are tested before cool-down.
The sample parameters, obtained via SEM visualization or transport measurement, are gathered in table \ref{parameters}.

The length of the normal metal, which varies between 0.75 and 1.9 $\mu m$, is much greater than the superconducting coherence length $\xi_S=\hbar D/\Delta$, where $\Delta$ is the BCS superconducting energy gap (see ratio $E_{Th}/\Delta$ in the table). Thus the devices are in the long junction regime, in which the properties of the normal metal  are known to determine the proximity effect, rather than the properties of the superconductor (\cite{Heikkila}).

The two superconductors used, Nb and Al, differ by their transition temperature. The fabrication procedure also produces different S/N interface conditions: In the Nb/Au samples, the superconductor is a homogeneous bilayer of Nb/Au and the normal metal is a bare Au region.
In contrast, in the angle-evaporated samples of Al/Au, the bare Au wires contact a region where Au and Al overlap (see SEM image). In addition, the Al loop is at some points, (depending on the loop shape, see picture) covered by Au. Therefore the superconductor in this configuration is less homogeneous.

The geometry of the SNS long junction is similar to previously measured ones (\cite{Dubos,Hoffmann}). The dcSNSquid geometry has to our knowledge not yet been reported in mesoscopic samples. A dcSNSquid with sub millimeter-wide junctions was however produced by Clarke and Paterson in 1971 \cite{Clarke}, in the aim of realizing an amplifier via a tuning of the dc squid asymetry. This will be detailed further on. The SNS dcsquid consists of a superconducting loop interrupted by two normal metal wires. In our setup, superconducting leads in a four wire configuration connect the loop. In comparison, the usual dc squid is formed by a superconducting loop interrupted by insulating tunnel barriers. A dc supercurrent flows through such a dcsquid, and its maximum value, the switching current, is known to be modulated periodically by a magnetic field, with a period of one flux quantum $\Phi_0=h/2e$ through the loop area.

\begin{table*}[tbp]

\begin{tabular}{|c|c|c|c|c|c|c|c|c|} \hline
                            & Nb750&Nb1200&Alw900&Alw1300&Alw1250&Alsq900 & Alsq1500 &Alsq1900\\\hline
geometry                    &SNS dc squid    & SNS dc squid  & SNS wire    & SNS wire  & SNS wire   & SNS dc squid  & SNS dc squid  & SNS dc squid \\\hline
$L_N (\mu m)$           &0.75      & 1.2    & 0.9     & 1.3      &  1.25   & 0.9     &  1.5    &  1.9   \\\hline
$W(\mu m)$             &0.4       & 0.4    & 0.125     & 0.125  &  0.125   & 0.13   &  0.15   &  0.2   \\\hline
$\Delta$                    &16 K      & 16 K   & 2.8 K   & 2.8 K    &  2.8 K  & 2.8 K   &  2.8 K  &  2.8 K  \\\hline
$E_{Th}$ from high T fit    &140 mK     & 50 mK  & 33 mK   & 36 mK    &    44 mK& 47 mK   &  30 mK  &  26 mK  \\\hline
$\Delta/E_{Th}$             &114       & 320    & 85     & 78        &  64     & 60      &  93     &  108  \\\hline
$\xi_S=L_N\sqrt{E_{Th}/\Delta}(nm)$&   70  & 67  &  100   &  150     & 156   & 116    &  155   & 180  \\\hline
$R_N$ from dV/dI ($\Omega$) &0.4       & 0.7    & 5      & 5         & 6       & 2.6 K   &  3.3    &  3   \\\hline
$I_s^{max} (\mu A)$                &330 & 68    & 13     & 3.7       & 4.3     & 17      &  7.6    &  4.6 \\\hline
$I_s^{min} (\mu A)$         &170 & 26    &  -    & -     &  -    &    9.2   & 1.8     & 0.6  \\\hline
$I_r (\mu A)$                       &75 & 16    & 1      & 0.6       & 0.6     & 1.3     &  1      &  0.45 \\\hline
b                           & 11      &  11&  22  &  6  &  7  &  11  &  10  &6  \\\hline
$I_s/I_r$										&4.4&  4.25 &  13  &  6.2 &   7.2& 13  &  7.6   &10.2  \\\hline
$\sqrt{2b}$									&  4.7     &  4.7  & 6.6  &  3.5  &  3.8  &  4.7  & 4.5  &  3.5\\\hline
$T^{*}$	(mK)										& 100 &  40  &  35  &  20  &  25  &  35  &  21  &  14\\\hline	

\end{tabular}

\caption{Characteristics of the different samples. $L_N$ and W are respectively the length and width of the normal metal, which was 50 nm-thick high purity ($99.9999 \%$) gold for all samples. $E_{Th}=\hbar D/L_N^2$ is the Thouless energy, with $D$ the diffusion constant. Note that $\xi_S=\sqrt{\hbar D/\Delta}$ is the superconducting coherence length determined from the gap of the superconductor but using the diffusion constant in the normal wire, so that it is evaluated via $\xi_S=L_N\sqrt{E_{Th}/\Delta}$.
The Al SNS wires (labeled Alwx, with x the length of the normal wire in nm) were produced on the same chip, and therefore should have the same material constants. This is also true for the Al SNS dc squids (labeled Alsqx), and for the Nb SNS dc squids (labeled Nbx).
The superconducting gap $\Delta$ is deduced from the transition temperature of the superconducting contacts, using $\Delta=1.76 T_c$}
\label{parameters}
\end{table*}

The measurements were conducted via room temperature $\pi$ filters, home made cryogenic lossy coaxial cables leading to the dilution stage, and 150 pF parallel capacitors on the sample holder. The samples were current-biased with a dc current up to $400 \mu\text{A}$ and a small ac current (a few hundred nA) at a few hundred Hertz for differential resistance measurements.

\subsection{Transition temperatures}

\begin{figure}
\begin{center} 
\includegraphics[clip=true,width=7cm]{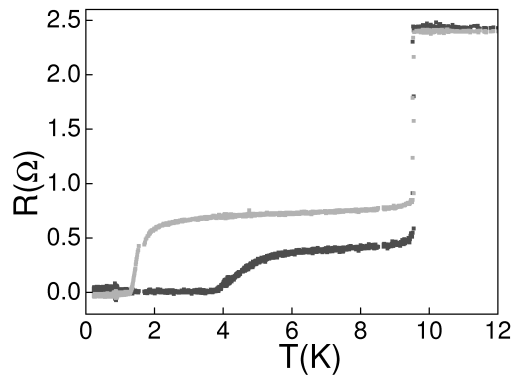}
\end{center}
\caption{Resistance of the two Nb/Au proximity dc squids versus temperature during cool-down. The configuration is a 4-wire measurement. The first transition around 9 K is the superconducting transition of Niobium. The transition at lower temperature (around 5 K for the shortest wire and 2 K for the longest one) is the transition of the gold wire from normal to a proximity-induced superconducting state.}
\label{Cool} 
\end{figure}

\begin{figure}
\begin{center} 
\includegraphics[clip=true,width=7 cm]{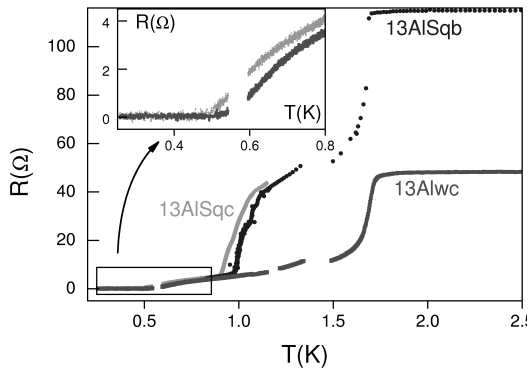}
\end{center}
\caption{Resistance of three Al/Au/Al wires as a function of temperature during cool-down. Sample Alw1250 is measured in a four wire configuration, Alw1300 and Alw900 were measured in a three wire configuration and a constant resistance of respectively 79 and 72 $\Omega$, corresponding to the resistance of one lead, has been subtracted. The double transition at 1.6 K and 1 K for samples Alw900 and Alw1300 is due to their geometry, in which part of the superconducting wire is made of pure Al with a critical temperature of 1.6 K, and part is made of an Al/Au bilayer, with a critical temperature of 1 K.} 
\label{CoolAl} 
\end{figure}

The resistance of two Au/Nb dc squids is plotted in Fig. \ref{Cool} as they are cooled to low temperature. The first resistance drop corresponds to the transition to a superconducting state of the Nb/Au bilayer, around 9.5 K. Below the transition temperature, the resistance varies by less than $10 \%$ over a large temperature range. The final resistance drop to a zero resistance state, the sign that the proximity effect develops through the entire normal metal so that a supercurrent flows through the normal metal, occurs at a lower temperature, roughly between $20$ and $40$ $E_{Th}$ depending on the samples. This temperature corresponds to the temperature at which the normal wire becomes phase coherent, and the proximity effect can set in. It is thus expected that it should correspond to $\min(L_T,L_\phi)=L/2$, i.e. $k_BT=4 E_{Th}$ in the limit where $L_T$ is smaller than $L_\phi$. The fact that the transition temperature is 5 to 10 times larger than this limit recalls the factors 3 and 10 in front of the Thouless energy in the expressions of the induced minigap or the critical current decay energy (see related paragraphs). The exact \``proximity effect transition temperature\'' has to our knowledge not been calculated theoretically. 

The behavior of the three Al/Au SNS wires, plotted in Fig. \ref{CoolAl} is similar.

\subsection{I V curves and differential resistance}

\begin{figure}
\begin{center} 
\includegraphics[clip=true,width=7cm]{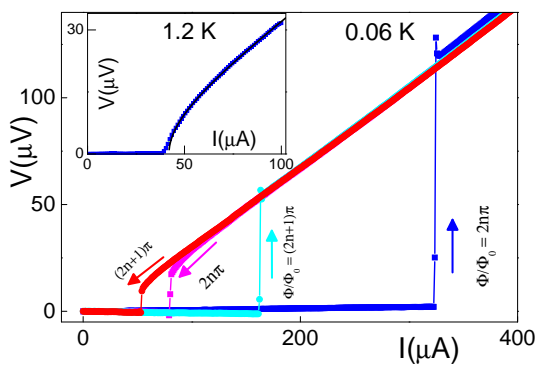}
\end{center}
\caption{Typical $V(I)$ curves of SNS dc squids at two magnetic fields, one in which the interference between both branches is constructive ($\Phi/\Phi_0=2n\pi$) and the other destructive ($\Phi/\Phi_0=(2n+1)\pi$), showing a maximum and minimum switching current. The hysteresis is clearly visible and characterized by a large ratio (roughly 3) between switching current $I_s$ and retrapping current $I_r$. The sample measured here is the shortest Au/Nb dc squid Nbsq750. The $V(I)$ curves for SNS wires are similar. Inset: IV curve of the same sample at 1.2 K,when there is no hysteresis. The shape of the curve corresponds to the square root dependence expected in the overdamped regime, $Q<1$ (continuous line).}
\label{VI} 
\end{figure}

 \begin{figure}
\begin{center} 
\includegraphics[clip=true,width=7cm]{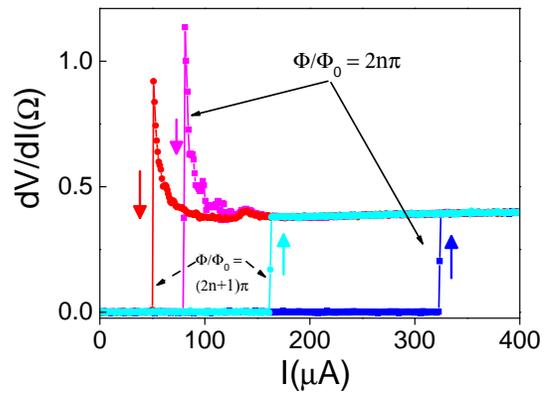}
\end{center}
\caption{Differential resistance of the shortest Au/Nb dcsquid (Nbsq750), at maximum and minimum switching current.}
\label{dVdI} 
\end{figure}

\subsubsection{Experimental curves at low temperature}
Typical $V(I)$ curves are shown in fig. \ref{VI} for a SNS dc squid at two magnetic fields, one in which the interference between both branches is constructive ($\Phi/\Phi_0=0$ or multiple of $2\pi$) and the other such that the interference is destructive ($\Phi/\Phi_0=(2n+1)\pi$). Single SNS junctions in the long limit have similar IV curves (not shown).
As the current is increased, a supercurrent flows (zero voltage drop), up to a value called the switching current (noted $I_s$), above which a voltage drop appears, and the junction is said to have switched to the dissipative branch. 
As the current is decreased, the switch from the dissipative branch to the superconducting one occurs at a current, called the retrapping current (noted $I_r$), which is much smaller than the switching current: the I V curves are strongly hysteretic, as found in previous experiments (\cite{Dubos,Anne,Hoffmann}).
The value of the retrapping current does not depend on how far we sweep the DC current, nor does it depend on the temperature, in stark contrast with the switching current, as described in the next section.

Equivalently, the differential resistance as a function of DC current is shown in fig. \ref{dVdI} for an SNS DC squid. The resistance in the normal state varies with current by a few percent in the Nb samples, and much more in the Al samples, probably because of inhomogeneities in the Al and Al/Au superconducting leads, which can generate phase slips. Some features on the down curve before retrapping (strongly fluctuating differential resistance just before retrapping, and resistance bump at 140 $\mu A$)are still not fully understood, but should be explained by the dynamics of an SNS junction out of equilibrium\cite{Argaman, Dubos}.

In the following paragraphs, we address the questions of the value of the switching current, which is 10 to 20$\%$ smaller than the critical current of the junction, the degree of hysteresis in the IV curves, and the fluctuation in the switching current for a given junction, field and temperature.
 
\subsubsection{Dynamics of the SNS junction: the RSJ model adapted to SNS junctions}

The IV curves are well understood in SIS tunnel junctions, which are modeled by a Josephson element in parallel with a capacitance and a resistive element (the so-called resistively shunted junction model, see for instance ref. \cite{Tinkham}). The elements of the circuit lead to a dynamical equation for the superconducting phase of the junction $\varphi$ given by:

\begin{equation}
\ddot{\varphi}+\frac{1}{\tau_R}\dot{\varphi}+\omega_p^2(sin(\varphi)-\frac{I}{I_c})=0
\end{equation}
Here $\omega_p=1/\sqrt{L_KC}$ where $L_K$ is the kinetic inductance given by $L_K=\phi_0/2\pi I_c$, and $\tau_R=RC$ controls the charge relaxation through the junction.
The state of the junction is represented by a particle whose position is the phase evolving in a tilted oscillating potential $U(\varphi)=-E_J(cos\varphi+I/I_c\varphi)$. The slope is the biasing current and the height of the barrier is related to the Josephson energy $E_J=(\hbar/e)I_c$. The time scale $\tau_R=RC$ represents friction for the particle. In the non dissipative state the particle is trapped in a local minimum, and oscillates at the plasma frequency $\omega_p/2\pi$. Increasing the current increases the slope of the potential and decreases the barrier height. Eventually the particle escapes at $I_s\leq I_c$.  The escape being a stochastic process, $I_s$ is characterized by a certain dispersion which is related to the thermal activation or the quantum tunneling of the particle through the barrier.
The hysteresis is also understood within this model. On the dissipative branch, when decreasing the current below $I_s$, the current consists in the externally imposed dc current plus an ac contribution due to the Josephson relation $\dot{\varphi}=eV/\hbar$.
At high dc voltages and currents, the ac part of the current, oscillating at the Josephson frequency $\omega_J=h/2eV$, flows entirely through the capacitance. The dc part of the current is dissipated in the resistance, and the IV curve is linear. At lower voltage, corresponding to Josephson frequencies approaching the circuit's resonant frequency $\omega_p$, the ac current starts to dissipate in the resistance, giving an additional contribution to the IV curve.
The condition $\omega_p=\omega_J$ thus corresponds to the onset of non linearities in  the IV curve and can be chosen as a criterion for the retrapping condition \cite{Tinkhamretrapping}, leading to $I_s/I_r\approx RC\omega_p=Q$. This explains that the retrapping current can be much smaller than the switching current if $\tau_R\gg1/\omega_p$ or $Q\gg1$.

\paragraph{Damping and hysteretic IV curves}

However this model is not directly applicable to the experiments on SNS junctions. Indeed the resistance and geometrical capacitance of SNS junctions (R a few Ohms \cite{Rqp}, and C roughly $10^{-16}$ fF) would correspond to $Q\ll 1$. In this so called overdamped regime, the expected I(V) dependence is $V=\sqrt{I^2-I_c^2}$, and therefore no sharp switching or hysteresis should be observed \cite{Tinkham}. Experimentally we find $I_s/I_r\approx Q$ between 3 and 7, see table I, and in other groups hysteresis is always observed at low temperature (Dubos {\it et al.} find 8, Anthore finds between 5 and 6 \cite{Anne}). A similar discrepancy has already been pointed out in superconducting weak links by Song \cite{Song}. The large hysteresis was explained by the fact that the pair relaxation time in the weak link was no longer given by $RC$ but by $h/\Delta$. Following this analysis, we replace the $RC$ time by the diffusion time of the Andreev pairs in the normal region $\tau_D=\hbar/E_{th}$. Comparison with the RSJ model is then achieved by replacing the capacitance with an effective capacitance $C_{eff}=\tau_D/R$. This yields an expression for the quality factor:  $Q=I_c/I_r=\sqrt{2e/hR_NI_c/\tau_D}=\sqrt{2eR_NI_c/E_{Th}}=\sqrt{2b}$. Table \ref{parameters} shows that the quality factor is indeed of the order of this value.

\begin{table}[htbp]

\begin{tabular}{|c||c|c|} \hline
   & SIS & SNS\\\hline
$eR_NI_c$&  $\Delta$ &$bE_{Th}$\\\hline
$E_J$ &$(\hbar/e)I_c$ & $(\hbar/e)I_c=\frac{b}{2\pi}\frac{R_Q}{R}E_{Th}$\\\hline
$'RC'$&RC & $\hbar/E_{Th}$\\\hline
$E_C$ &$e^2/C$ &$2\pi \frac{R}{R_Q}E_{Th}$ \\\hline
$\hbar\omega_p$&$\sqrt{\hbar2eI_c/C}$ &$\sqrt{2b}E_{Th}$ \\\hline
Q&$\omega_p RC$ & $\sqrt{2b}$\\\hline
$k_BT^*$&$\hbar\omega_p/(2\pi)$ &$\frac{\sqrt{2b}}{2\pi}E_{Th}$ \\\hline

\end{tabular}

\caption{Comparison between quantities determining the switching process, for SIS and SNS junctions. Here $R_Q=h/e^2$ and $b=eR_NI_c/E_{Th}$.}
\label{compSISSNS}
\end{table}

Note that heating of the sample once the SNS junction has switched to the dissipative state will also cause hysteretic I V curves: The power injected in the normal wire may not be entirely dissipated by the substrate and superconducting contact phonons at low temperature, and heat conduction by the superconducting contacts is weak because of the superconducting gap. It is therefore probable that the electron temperature as the current is ramped down after switching is higher than before switching, so that the experimental retrapping current may be smaller than the intrinsic retrapping current.

Since both heating and intrinsic charge relaxation depend on the wire length (albeit with different power laws), it is difficult at this stage to determine exactly the relative contribution of each effect to the hysteresis. 
However the very shape of the IV curve  is a clear indication that these SNS junctions are intrinsically hysteretic, since a non-hysteretic (overdamped) junction would have a square-root dependence $V=R\sqrt{I^2-I_c^2}$ of the V(I) curve \cite{Tinkham}, in contrast with the sharp jump observed at the transition in our junctions.
Furthermore, measurements of the noise spectrum of long diffusive SNS junctions by Hoffmann {\it et al.} \cite{Hoffmann} suggest that the electron temperature is less than 500 mK, even in the retrapping branch.

It is therefore safe to conclude that although electron heating may contribute to the hysteresis in the IV curve, intrinsic hysteresis is also present and most likely due to the relatively slow dynamics of the phase and charge in the long normal wire.

The inset of fig. \ref{VI} shows that the hysteresis disappears at higher temperature. This could be due to a lower $I_c$ and to a more efficient heat dissipation at higher temperature, and therefore decreased heating. But it may also be explained by the fact that the quality factor Q depends on temperature via the square root dependence of the plasma frequency $\omega_p$ upon $I_c$ (see table II). In the case of this sample Nb750, the switching current is divided by 8 between low temperature and 1 K, and the quality factor should thus loose a factor $\sqrt{8}\approx 3$, bringing the quality factor to 1 at 1K, corresponding to strong damping and a non hysteretic curves with square root V(I) dependence, as is observed (see Fig \ref{VI}).

%
%The electronic temperature Tel corresponds no more to the phonon bath temperature, and can be roughly estimated, if we assume that all the power injected contributes to heating the electrons, with the formula P=Sigma V (Tel^5-Tph^5). Explaining the hysteresis as an heating effect, we can suppose that Ir=Ic(Tel). Comparing Tel to the temperature T when Ir=Ic(T), we find, for all the samples, that Tel approx 2.4*T. Because of the uncertainty on the normal state resistance and of the simplistic model, it is difficult to distinguish between heating and intrinsic charge relaxation, even if they depend on the wire length with two different power laws.
%

\paragraph{Fluctuations of switching current}

\begin{figure}
%\begin{center} 
\includegraphics[clip=true,width=9 cm]{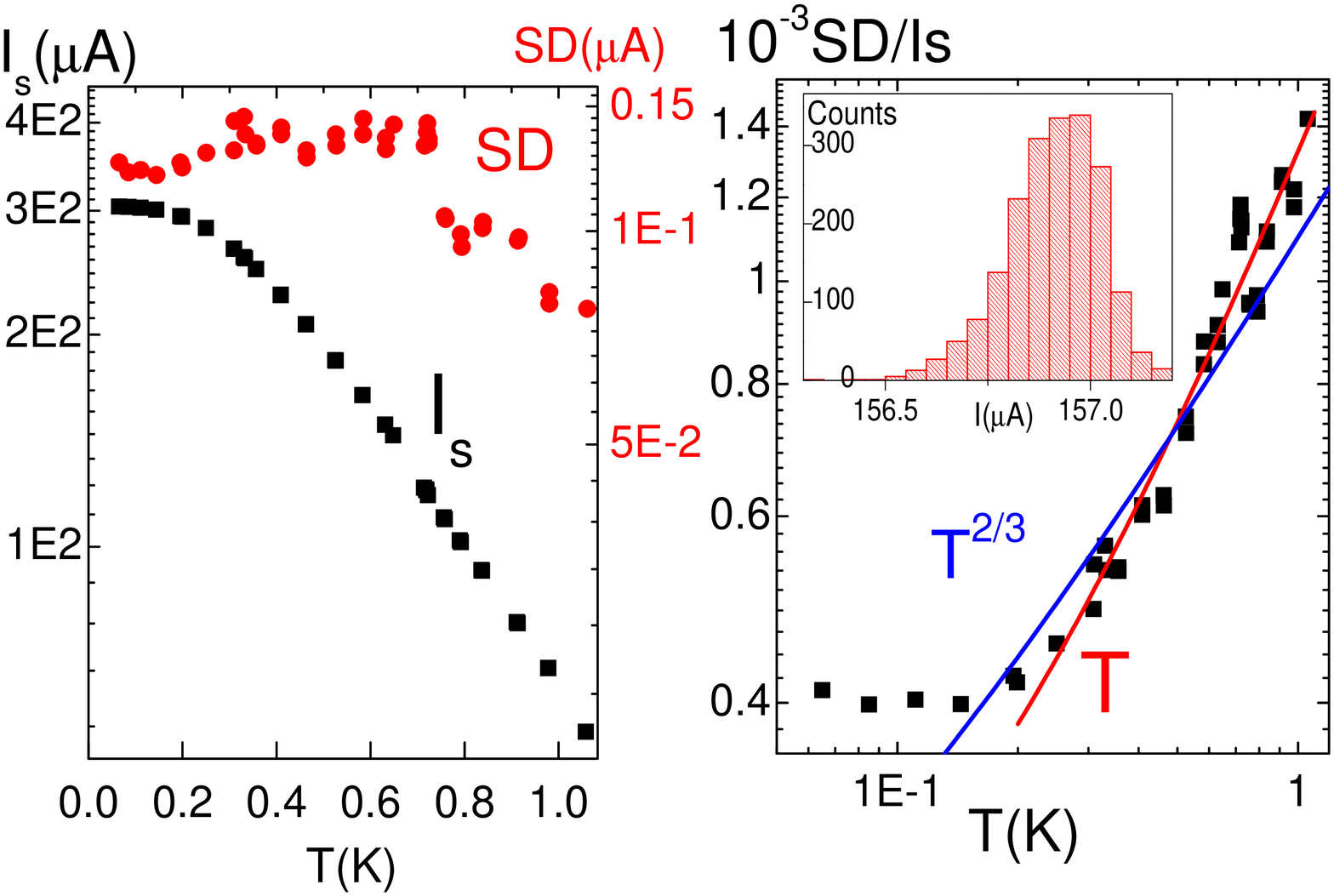}
%\end{center}
\caption{Left pannel: Switching current $I_s$ and width of the histogram (Standard deviation $SD$) as a function of temperature for the short Nb dc squid (Nbsq750), at $\Phi = 0$. Right pannel: ratio $SD/I_s$, compared to a linear and $T^{2/3}$ dependence. Inset: Histogram of the switching current for $\Phi/\Phi_0 = (2n+1)\pi $ for the shortest junction Nb/Au ring. The total number of counts is 2000.
}
\label{Histo_T} 
\end{figure}

The dynamics of the SNS junction also determine the probability distribution of the current values at which the junction switches from its superconducting to normal state. We have measured histograms of the switching current, following the technique used by Devoret {\it et al.} \cite{Devoret}. The current is repeatedly swept (at a frequency of $f=65~Hz$) from a small negative current to $I_{\text{max}} > I_c$, and the current at which a finite voltage appears is recorded. Fig.\ref{Histo_T} shows a histogram of the switching current for a Nb/Au ring at $\Phi/\Phi_0=(2n+1)\pi$. The histograms of Al/Au junctions are similar. As expected from a switching phenomenon, the histogram is asymmetric, falling of more sharply at the higher currents. The histogram is narrow, with $\delta I_s/I_s$ roughly 1 per thousand.

Following the previous description of the hysteresis, in which the geometrical capacitance C is replaced by an effective one $\tau_D/R$, we can extrapolate the SIS theory to the SNS case. All the characteristic scales for SIS junctions and their correspondence for SNS junctions are listed in table \ref{compSISSNS}.
Garg \cite{Garg} has calculated the width of the switching histogram of SIS junctions. This width depends on the damping and the origin of the escape rate (thermal or quantum). In the underdamped regime, which is our situation (see table), at high temperature the particle should escape by thermal activation and the width is given by $\delta I_s/I_s\approx (T/E_J)^{2/3}$ but below a temperature $T^{*}=\hbar\omega_p/2\pi k_B$, of the order of $E_{Th}$, the particle escapes by quantum tunneling and the width of the histogram is given by $\delta I_s/I_s\approx (E_{Th}/E_J)^{4/5}$. Extrapolating these results to the SNS junction, the expected width is proportional to $(T/E_J)^{2/3}=(2\pi R_N/R_Q.T/E_{Th})^{2/3}$ in the thermally activated regime, and proportional to $(E_{Th}/E_J)^{4/5}=(2\pi R_N/R_Q)^{4/5}$ in the quantum regime. The predicted widths are therefore of the order of $\delta I_s/I_s=$ 1 per thousand, as measured experimentally at low temperature. The 2/3 power law however in the thermally activated regime is not really found experimentally (see Fig. \ref{Histo_T}).

%\begin{figure}
%\begin{center} 
%\includegraphics[clip=true,width=7cm]{Histo.eps}
%\end{center}
%\caption{Histogram of the switching current for $\Phi = \pi$ for the shortest junction Nb/Au ring. 
%The total number of counts is 2000.
%}
%\label{Histo} 
%\end{figure}

\subsection{Temperature dependence, Thouless energy and normal state resistance}

The critical current is predicted \cite{Heikkila,WilhelmZaikin} to reach $eR_NI_c(T=0) = 10.82~E_{Th}$ in the limit of infinitely long junctions ($\Delta/E_{Th}=\infty$), and to be given by $eR_NI_c(T=0) = b E_{Th}$, with b smaller than 10.82 but within a few percent as soon as $\Delta/E_{Th}>10$. The ratios $b=eR_NI_s/E_{Th}$ at the lowest temperature (16 mK for Nbsq and Alsq samples, 60 mK for Alw samples) are listed in table \ref{parameters}  and are within $50\%$ of this value. The main uncertainty in calculating the ratio however is in determining the normal state resistance. The Thouless energy $E_{Th}$ is extracted with greater confidence from the temperature dependence of the switching current, as explained in the next paragraph.

\begin{figure}
%\begin{center} 
\includegraphics[clip=true,width=8 cm]{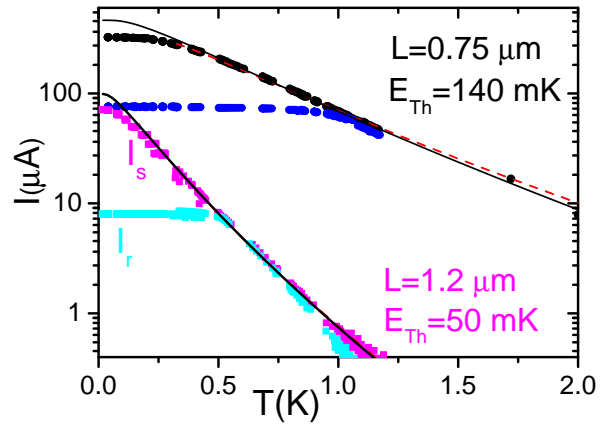}
%\end{center}
\caption{Switching and retrapping current as a function of temperature for both Nb dc squids, at maximal switching current, comparison between experiment and theory. Filled dots and squares are the data points. Red dashed lines are fits to the high temperature formula for the proximity effect (see Eq (1) of Dubos {\it et al.}, from which the Thouless energy is deduced. Black lines are the numerical solutions to the Usadel equations using the Thouless energy deduced from the high temperature fit.
The hysteresis disappears (i.e. the switching and retrapping current are equal) at a temperature close to $11 E_{Th}$.\\
\label{IcIr_T}}
\end{figure}

The switching current can thus be defined (rather equivalently) as either the average or the most probable value out of a series of measurements, and is plotted for 2000 measurements as a function of temperature in Fig. \ref{IcIr_T}. As is clear in the figure, the decay of switching current above a few hundred milliKelvin is roughly exponential:  $I_s\approx e^{-T/\alpha E_{Th}}$, with $\alpha$ of order 9 \cite{WilhelmZaikin}. We used an approximate high temperature expansion (Equation (1) of Dubos et al. \cite{Dubos}, valid when $k_B T > 5 E_{Th}$) to determine the Thouless energy and the normal state resistance (dashed lines). The full temperature dependence, obtained by numerical integration of the Usadel equations using the determined Thouless Energy, is also plotted as a continuous line of the figure. For both Al and Nb samples the high temperature decay of the switching current is well described by the Usadel calculation. At low temperature however, the predicted switching current is larger than found experimentally (by roughly a factor two, for instance 100 $\mu A$ instead of 70 $\mu A$ for the longests Nb dc squid (Nbsq1200), and 500 $\mu A$ instead of 250 $\mu A$ for the shortest dcsquid (Nbsq750)). The deviations from the theoretical prediction occur below 400 mK for the longest junction, and below 200 mK for the shortest one. The switching current saturates at lower temperatures. We have no explanation for the deviations from theory at temperatures which are relatively high: although the electronic temperature may be higher than the phonon temperature of the mixing chamber at the lowest temperatures (16 to 50 or 100 mK), it is improbable that the electron temperature be very different than the phonon temperature at temperatures higher than 100 mK \cite{temperature}.

The fitting parameters $R_N$ and $E_{Th}$ are given in Table 1, as are the value of the ratio $eR_NI_s/E_{Th}$.
The normal state resistance $R_N$ extracted from the fit is somewhat ($50\%$) smaller than that extracted from the dV/dI curves, and the $eR_NI_s/E_{Th}$ product is also roughly $50\%$ less than maximum value 10.82, expected at zero temperature, for an infinitely long normal wire and a perfect transparency.

The role of non perfect transparency of the SN interface has been investigated in ref \cite{Cuevas}. Numerical integration of the Usadel equations shows that the main effect of a non transparent interface is to renormalize the Thouless energy, and to decrease the region of current saturation: thus an imperfect interface could not explain the slower variations with temperature of the switching current at low temperature.

%We also show in Fig. \ref{Histo_T} the behavior of the histogram width, for the shortest Nb sample, as compared to the switching current variations: whereas the switching current decays practically exponentially, the histogram width practically does not change with temperature, at roughly $0.5\%$ of $I_S$. 

%\begin{figure}
%\begin{center} 
%\includegraphics[clip=true,width=7 cm]{13Al_Is_T.eps}
%\end{center}
%\caption{Switching current as a function of temperature for Al SNS wires.} 
%\label{13Al_Is_T} 
%\end{figure}

%\begin{figure}
%\begin{center} 
%\includegraphics[clip=true,width=9 cm]{17Al_Is_T.eps}
%\end{center}
%\caption{Switching current as a function of temperature for 17Al SNS dcsquids.} 
%\label{17Al_Is_T} 
%\end{figure}

\subsection{Magnetic field dependence}

The variations of the switching current in a perpendicular magnetic field is central to our investigation.
As for conventional SIS dc squids, the SNS dc squid samples have a periodically oscillating switching current, whose period corresponds to a flux quantum $\Phi_0=h/2e$ through the loop area, as shown in figures \ref{Al17ch_th},\ref{asymfillong},\ref{asym},\ref{17AlbIs_H}, and \ref{IcHNbcourt}. There are however features of the field dependence that have not been reported before, for instance the absence of the Fraunhoffer pattern in the decrease of switching current at larger field: the decay is monotonous, practically gaussian with a field scale of one flux quantum through the normal wire.

\subsubsection{Modulation of the switching current by the magnetic field: the DC SNSquid}

%\begin{figure}
%\begin{center} 
%\includegraphics[clip=true,width=7cm]{Histo_Hcourt.eps}
%\end{center}
%\caption{Critical current of the short junction Nb/Au squid and width of the histogram $SD$ (Standard deviation) for one period.}
%\label{Histo_Hcourt} 
%\end{figure}

\begin{figure}
\begin{center} 
\includegraphics[clip=true,width=7cm]{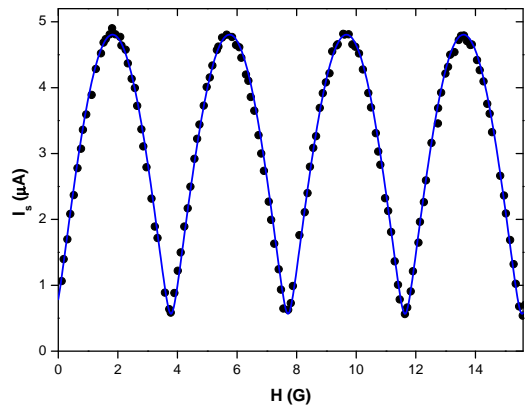}
\end{center}
\caption{Switching current of Alsq1900 SNS dc squid as a function of magnetic field at T= 16 mK (circles), and fit to Equation \ref{squid} (line) showing the dc squid-like behavior. Here $I_1=2.1~\mu A$ and $I_2=2.7~ \mu A$.}
\label{Al17ch_th} 
\end{figure}

\begin{figure}
\begin{center} 
\includegraphics[clip=true,width=7cm]{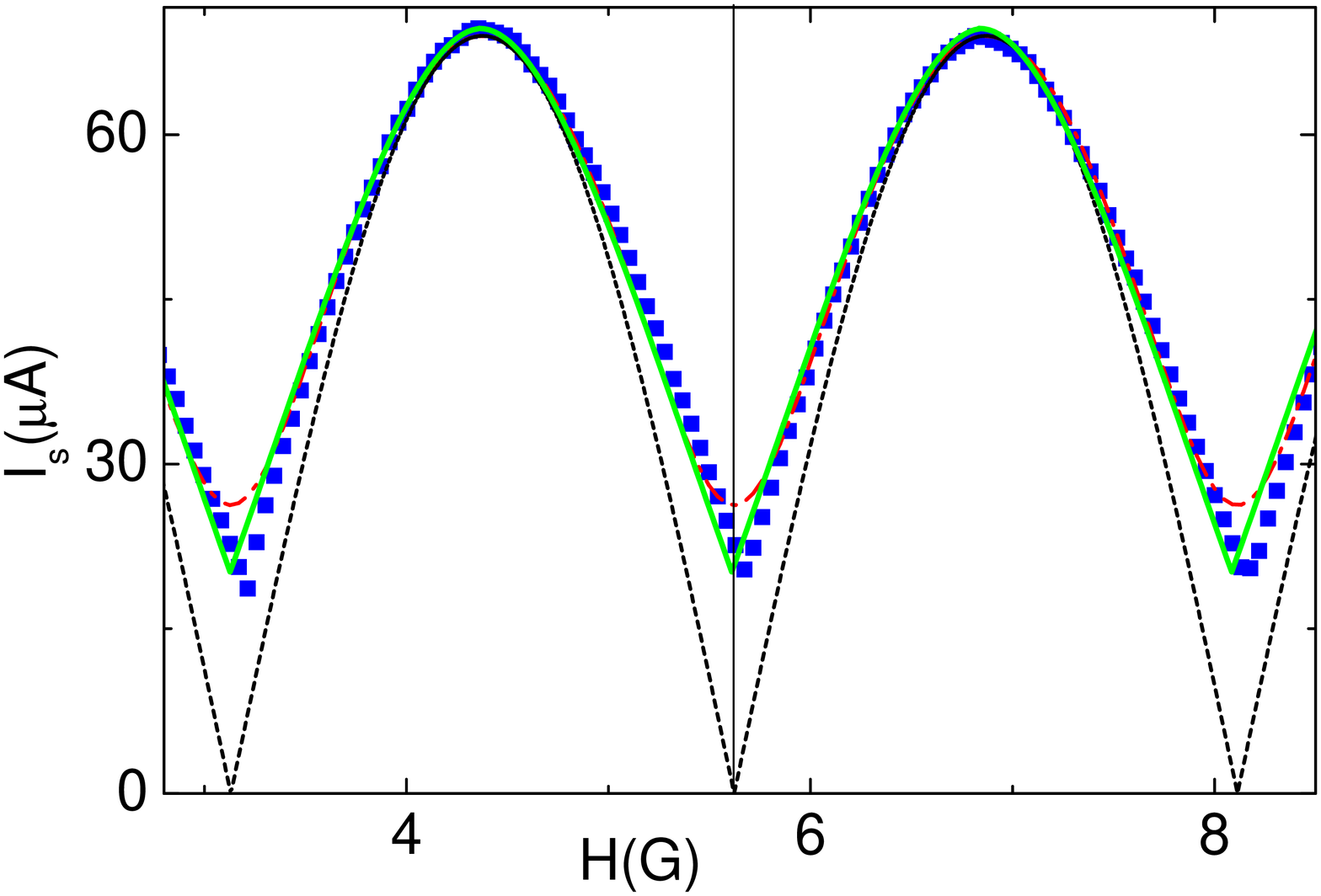}
\end{center}
\caption{Dots: Switching current over three magnetic field periods, for the longest Nb dc squid Nbsq1200, measured at 60 mK. Lines are the simulated curves corresponding to the same maximum switching current of 70 $\mu A$: black dotted curve, a non inductive squid with identical junctions (same critical current); red dash-dotted curve, a non inductive squid with different junctions; and green line, a squid with identical junctions but two inductive branches, with identical inductance L=6 pH.}
\label{asymfillong} 
\end{figure}

\begin{figure}
\begin{center} 
\includegraphics[clip=true,width=8cm]{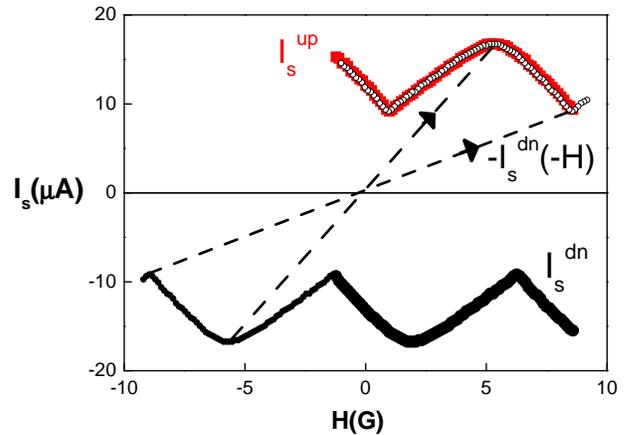}
\end{center}
\caption{Sample Alsq900, comparison between the positive switching current in the up sweep (increasing current, red square symbols) and the negative switching current in the down sweep (decreasing current, black round sumbols). The black curve with small open dots is the transformation of the down switching curve, via the  time reversal operation $T(I_s^{dn}(H))= -I_s^{dn}(-H)$, and is seen to be superimposed on $I_s^{up}(H)$, as expected from the  global time reversal symmetry of the experiment.}
\label{asym} 
\end{figure}

\begin{figure}
\begin{center} 
\includegraphics[clip=true,width=9 cm]{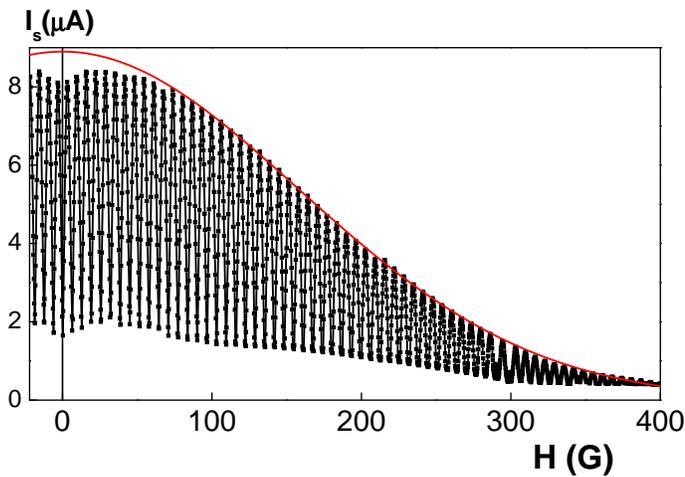}
\end{center}
\caption{Switching current as a function of magnetic field for Al SNS dc squid Alsq1500, at 17 mK. The line is a Gaussian.} 
\label{17AlbIs_H} 
\end{figure}

\begin{figure}
\begin{center} 
\includegraphics[clip=true,width=9cm]{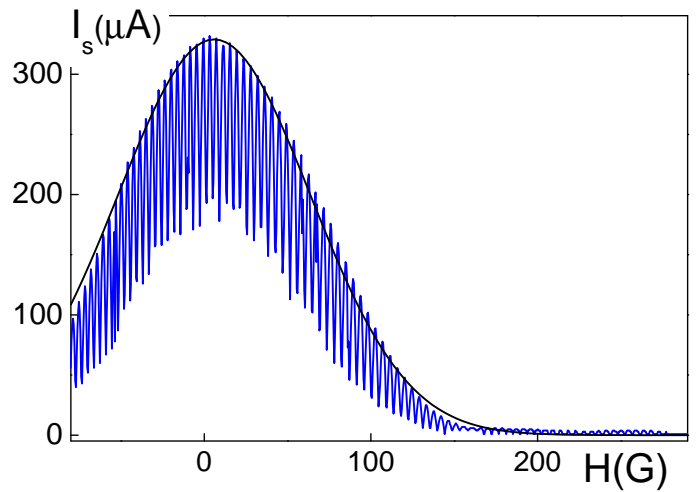}
\end{center}
\caption{Switching current of the short junction Nb/Au dc squid Nbsq750 as a function of magnetic field, at 60 mK. Black line is a Gaussian.}
\label{IcHNbcourt} 
\end{figure}

We first focus on the proximity SQUID interference patterns, presented in figures \ref{Al17ch_th} and \ref{asymfillong}, and compare them to the theory of a regular DC SQUID made of two tunnel SIS junctions in parallel \cite{Tinkham}. The critical current $I_c$, which is the switching current in the absence of fluctuations, is given by:
\begin{eqnarray}
I_c^2 = (I_1 -I_2)^2 + 4I_1I_2\cos^2(\pi \frac{\Phi}{\Phi_0})\\ = I_1^2+I_2^2 + 2I_1I_2\cos(2\pi \frac{\Phi}{\Phi_0}),
%\nonumber
\label{squid}
\end{eqnarray}
with $I_1$ and $I_2$ the critical currents of the 2 junctions, and $\Phi$ the magnetic flux threading the SQUID loop. This critical current thus oscillates periodically with field with a period of one flux quantum  $\Phi_0=h/2e$ through the squid loop. The oscillation amplitude is maximum in symmetric squids, and is reduced in  squids whose junctions have different critical currents. The relative amplitude in non inductive squids is given by 
\begin{equation}
r_{as}=\frac{I_c^{max}-I_c^{min}}{I_c^{max}}=\frac{2\min(I_1,I_2)}{I_1 +I_2}.
\end{equation}
Such a dc squid behavior is indeed demonstrated in our SNS dc squids, as shown in Figs. \ref{Al17ch_th}, \ref{asymfillong}, \ref{asym}, \ref{17AlbIs_H}, \ref{IcHNbcourt}.  Supercurrents up to 330 $\mu A$ are modulated by 45 to $88\%$, see table 1. The incomplete modulation, which corresponds to a ratio $I_1/I_2$ between $1.3$ and $3.4$, can be attributed to slight differences in the geometry of the two normal wires, to which the critical current is very sensitive. Indeed, the critical current of a long SNS junction scales as the inverse cube of the normal wire length.

The inductances $L_1$ and $L_2$ of the 2 branches of the SQUID also come into play in determining the shape of the modulation curve, since they act to screen the external flux $\Phi_{ext}$ by $\delta \Phi = L_1i_1 -L_2i_2$ where $i_1$ and $i_2$ are the currents through the 2 branches of the SQUID (and $i_1 +i_2 =I$).  In symmetric squids with $L_1 = L_2$, the screening induced is zero for fields corresponding to the maximum switching current, since the flux induced by the screening current in one branch is exactly canceled by the flux induced by the same current in the other branch. The maximum effect of the screening flux produced by the inductance corresponds to half a flux quantum in the loop
$\Phi_{ext}/\Phi_0 \equiv 1/2~mod~1$. This explains deviations from the simple cosine dependence seen in some samples (see for instance Fig. \ref {asymfillong} ). In some  cases, corresponding to  $L_1 \neq L_2$,  asymmetries with respect to reversal of magnetic field or current are clearly visible (see Fig. \ref{asym}).   The switching currents $I_s^{up}$ and $I_s^{dn}$ measured  on the positive and negative  current branches of $V(I)$ (while increasing the absolute value of current in both cases) differ at finite magnetic field:
\begin{eqnarray}
I_s^{up}(-H)\neq I_s^{up}(H)\\
I_s^{dn}(-H)\neq I_s^{dn}(H).
\nonumber
\end{eqnarray}

However, as expected from the global time reversal symmetry of the experiment, $I_s^{up}(H)= -I_s^{dn}(-H)$. 
These asymmetries have previously been calculated in \cite{Barone}, and measured in $S/I/S$ dcSQUIDS \cite {Ballestro}. They were also exploited by Clarke and Paterson \cite{Clarke} in larger area SNS dc squids: the authors controllably adjusted the asymmetry of the two superconducting branches and verified the corresponding asymmetric $I_s(\Phi)$ curves. The current modulation in those SNS dc squids was less than 5 percent.
As expected for self inductance effects, the asymmetry and anharmonic distortion of the periodic oscillations are largest in samples with the largest critical current. 

One also expects the intrinsic non harmonicity of the current-phase relation of an SNS junction to cause a distortion in the switching current Vs field curve of an SNS dc squid: the amplitude of the $m$-th harmonic in the $\phi_0$ periodic current phase relation is predicted to vary like $1/m^2$ \cite{Heikkila}
and should therefore be detected at least up to the 3rd order. However as is shown in the Appendix, the critical current of a dc SNSQUID with  2 symmetrical junctions is insensitive to even harmonics, so that the harmonics content of the current-phase relation does not easily show up.
To specifically test the anharmonicity of the current-phase relation in a long SNS junction, a strongly  asymmetric  setup  with one reference SIS harmonic junction  like the quantronium  \cite{atomiccontacts}, would in principle be more adequate. However, it would still be sensitive to self inductance effects for large supercurrents. Therefore the best way to measure the various harmonics of the current phase relation of long  metallic junctions is to directly measure the magnetic orbital response of a simple SNS ring, as was done  recently using a magnetic Hall probe technique \cite{Strunk}.

\subsubsection{Decrease of the critical current at high field}

We now discuss the decrease of the switching current at large field scales. In single junction devices, the extinction of the switching current with field is straightforward. In the dcSNSquid samples, whose switching current oscillates on a small field scale, we consider the extinction at large field of the envelope of the oscillations. Where necessary, we follow the behavior with field of the maxima of the oscillations. We will in the following neglect the effect of the asymmetries in the squids, and consider that the behavior in field of these maxima is equivalent to the behavior in field of a single SNS junction. 

The first observation is that we do not observe the Fraunhoffer diffraction pattern encountered in many experiments on wider and shorter SIS and SNS junctions \cite{Jaklevic,Finnemore,graphene}. Instead, we find a monotonous decay with a typical field scale of a flux quantum through the normal metal wire. For most of our samples, the decay is approximately gaussian.
 
Qualitatively, in a semi-classical picture, the decay can be seen as due to dephasing induced by the magnetic field: different diffusive Andreev paths in the normal metal acquire different phases. If the junction is wide, the phase along the NS boundary also varies, but in the limit of a normal wire much longer than it is wide, only the phase acquired during the propagation across the normal metal is relevant.

Quantitatively, the average of the phase factor $\phi(C) = \int_C \vec A \vec dl /\Phi_0$ acquired along a path C has been determined analytically in the 1D limit (normal wire length $L$ much greater than width $W$), using the solution of the field dependent diffusion equation \cite{Gilles07}. This leads to 
\begin{equation}
I_c(H)= I_c(0) \frac{\frac{\pi}{\sqrt{3}}\frac{H}{H_0}}{\sinh \frac{\pi}{\sqrt{3}}\frac{H}{H_0}}
\label{Gilles}
\end{equation}
where $H_0  = \Phi_0/WL$ corresponds to a flux quantum through the normal part of the SNS junction (and $\Phi_0=h/(2e)$). 

This calculation does not take into account the modification of the density of states in the normal wire. These experiments have also stimulated a full treatment of the proximity effect via the numerical resolution of the Usadel equations, for any ratio of the wire length versus width. The magnetic field enters in these equations via a depairing time $\tau_H= 6 \hbar ^2e^2/(e^2Dw^2H^2).$ Cuevas {\it et al.} \cite{Cuevas} have found that
in the 1D limit ($L\gg W$), the solution can be fit by a gaussian dependence at low fiels ($H/H_0<3$).

\begin{equation}
I_c(H)= I_c(0)\exp- \left[a(H/H_0)^2 \right] .
\label{Cuevas}
\end{equation}
the coefficient $a$ is worth 0.24 at zero temperature, and decreases as T increases.

For low fields ($H/H_0<2$), the semiclassical calculation equation \ref{Gilles} is also well approximated by a Gaussian, but the coefficient in the exponent is 0.55 instead of 0.24.

In SNS junctions with wider wires ($W>L$), Cuevas {\it et al.} \cite{Cuevas} find that the critical current does not decay monotonously with field but that oscillations appear, and that for very large junctions the well known Fraunhoffer pattern is recovered.

\begin{figure}
\begin{center} 
\includegraphics[clip=true,width=9 cm]{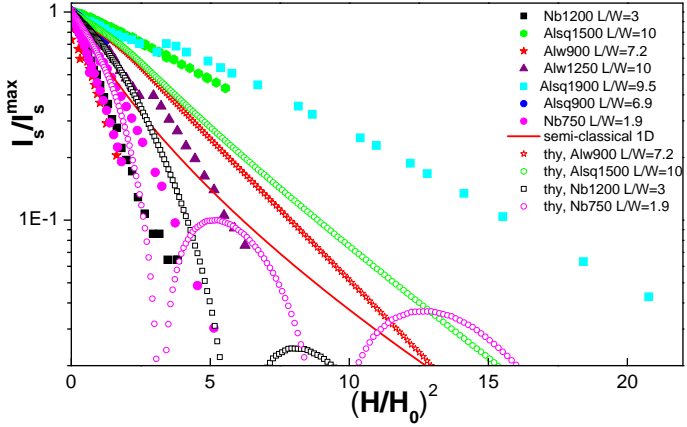}
\end{center}
\caption{Field dependence of the normalized switching current for all samples at lowest temperatures. For the samples with the squid geometry, the filled data points consist of the local maxima of the full (modulated) curves.
Also plotted with open symbols are the theoretical predictions of the Usadel equations at zero temperature, for four of the samples taking into account the exact aspect ratio of the junction. The prediction of the 1D semiclassical calculation Eq. \ref{Gilles} is the continuous line. The vertical scale is logarithmic, and the horizontal axis is the square of magnetic field normalized by the field $H_0$ corresponding to a flux quantum through the wire surface.} 
\label{AllIc_H} 
\end{figure}

The ratio $L/W$ of our samples  is respectively 2 and 3 for the short and long Nb samples, and exceeds 7 for all Al samples (see table \ref{parameters}). To compare experiment and theory, we thus computed the predicted field dependence of the maximal switching current, using the Usadel equations with the experimental aspect ratio of the junctions. The curves for both Nb SNS dc squids and two Al SNS samples are plotted in Fig. \ref{AllIc_H}.
It can be seen that the decay of the switching current is only qualitatively described by theory. In particular theory predicts that the widest Nb/Au/Nb junction should exhibit relatively large oscillations of the maximum switching current with field, which is not found in the experiment.

One possible explanation for these discrepancies is the screening of the magnetic field by the superconducting  contacts, as well as by the normal parts. Screening by the superconductor focuses the magnetic flux lines through the normal, leading to a smaller $H_0$. Screening by the normal metal on the other hand tends toward flux expulsion and a larger $H_0$. We have not yet included these effects in the calculations.

\begin{figure}
\begin{center} 
\includegraphics[clip=true,width=8 cm]{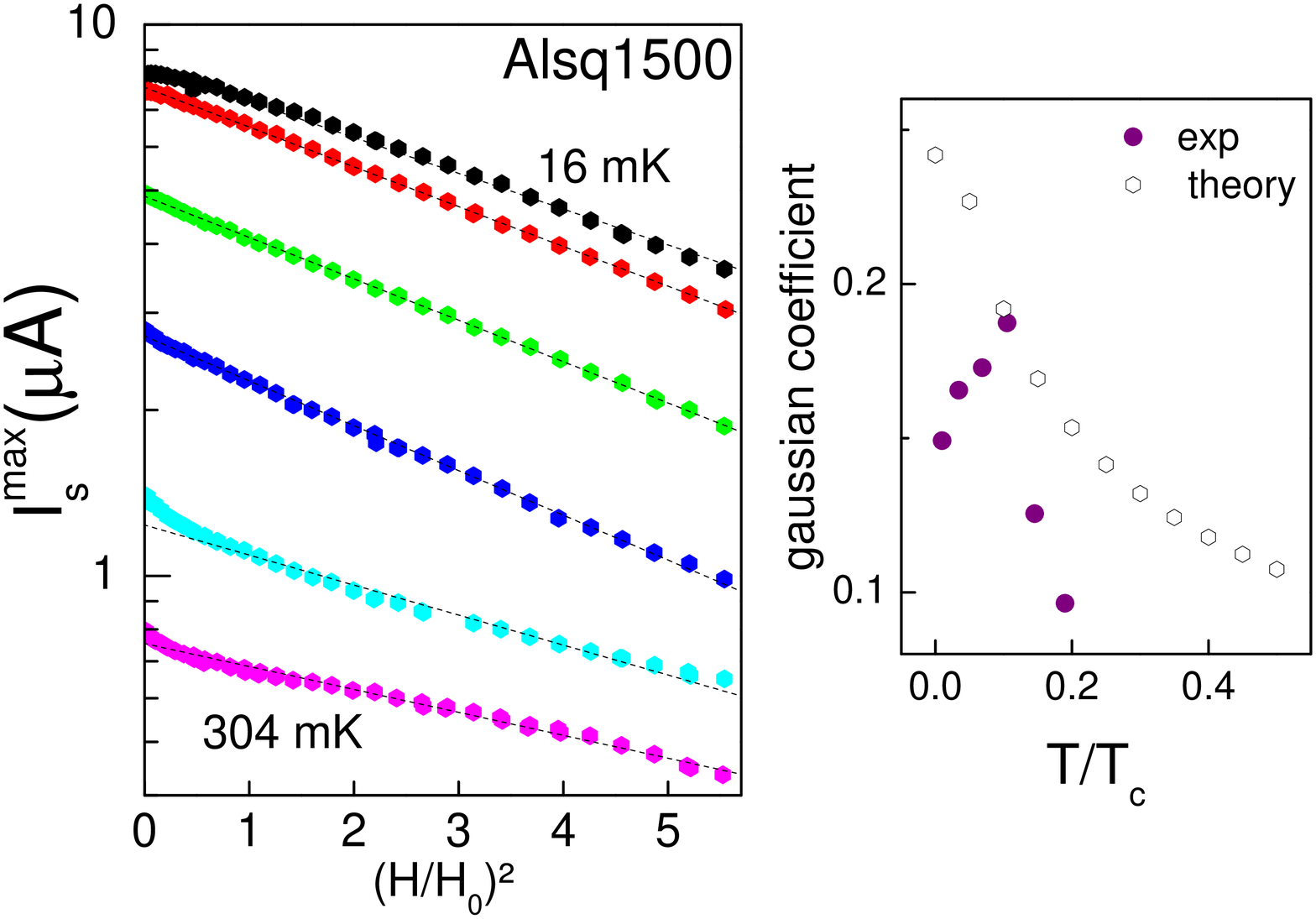}
\end{center}
\caption{Field dependence of the switching current, as a function of temperature. Left panel:Maximum switching current (extracted from the modulated curves) as a function of magnetic field for Al SNS dc squid Alsq1500, at temperatures of (from top to bottom) 16, 55, 110, 168, 233 and 304 mK (symbols).
The vertical scale is logarithmic, and the horizontal axis is the square of the reduced magnetic field. Dashed lines correspond to a Gaussian decay. Right panel: Full symbols, temperature dependence of the coefficient of the Gaussian decay (coefficient a of Eq. \ref{Cuevas}) extracted from the data, compared to the prediction of the Usadel equations (empty symbols).} 
\label{17AlbIs_H_T} 
\end{figure}

Fig. \ref{17AlbIs_H_T} shows the field dependence of sample Alsq1500 at several temperatures, the temperature dependence of the Gaussian coefficient, and the comparison with the coefficients predicted by the Usadel equations. Here also experiment and theory differ, a fact that may be due to the temperature dependence of screening of the magnetic field.

%\begin{figure}
%\begin{center} 
%\includegraphics[clip=true,width=8cm]{IcHNblong+fits.eps}
%\end{center}
%\caption{Switching current of the long junction Nb/Au dc squid as a function of magnetic field. Red dashed curve %is formula \ref{Gilles}, and black line is formula \ref{Cuevas}.}
%\label{IcHNblong} 
%\end{figure}

\subsubsection{Reentrant behavior at low field and low temperature}

\begin{figure}
\begin{center} 
\includegraphics[clip=true,width=8 cm]{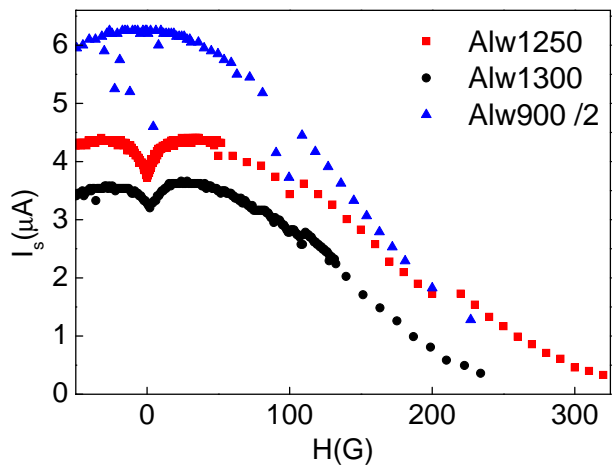}
\end{center}
\caption{Switching current as a function of magnetic field for the three Al/Au SNS wires, at 17 mK. Switching current of wire Alw900 has been divided by 2 for better comparison with the two other samples. Two of the three wires show a reentrant behavior.} 
\label{13AlIs_H} 
\end{figure}

Certain samples exhibit a surprising increase  of  the switching current at low magnetic field (below $20 G$, see  fig.\ref{13AlIs_H}) and at temperatures below 150 mK. A qualitatively similar effect  but on a very different field scale, was already observed on narrow superconducting wires made of MoGe and Ge \cite{Bezryadin} and explained by phase breaking of Cooper pairs induced by spin flip scattering on magnetic impurities. Such spin flip scattering disappears when the magnetic moments are polarized under a magnetic field such that $\mu_B B = k_B T$. This would correspond to a field of a few hundred Gauss in the present experiment, more than an order of magnitude larger than observed.
Similar anomalies at low field were also reported by Dynes {\it et al.} \cite{Dynes} in narrow supercondcuting wires, and attributed to phase fluctuations of the superconducting order parameter.
Other possible explanations include:

- Flux trapping in the superconducting wires during cooling of the samples. This can probably be excluded since no irreversibility or hysteresis is associated  to this unusual field dependence. 

- Quantum interference in the normal part of the junction. They are known to lead to weak localisation corrections of the conductance in the absence of proximity effect. The correction is of the order of $e^2/h$, which for a wire of roughly $1~\Omega^{-1}$ corresponds to $\frac{\delta i}{i}\approx 10^{- 4}$, i.e. much less than the 10 or 20$\%$ observed in the experiment. Moreover, the reentrance  disappears around the temperature corresponding to the Thouless  energy, to which weak localisation is not sensitive.

- A last possibility could be a magnetic orbital effect, in the form of quasiparticle persistent current loops, giving rise to strong paramagnetic orbital contributions similar to what was observed in hybrid N/S cylinders at very low temperatures \cite{Mota,Imry}. 

It is noteworthy that this reentrance of $I_s(H)$ is only present in the
aluminum S/N/S junctions  or SQUIDs, which have an abrupt interface between the normal and superconducting regions, and was never observed in the Nb junctions, in which the Au is also present under the Nb electrodes.  It also occurs only in normal junctions whose length is between 1.2 and 1.5 $\mu m$, and was not detected in shorter or longer wires.

%\begin{figure}
%\begin{center} 
%\includegraphics[clip=true,width=7 cm]{17Alah_Is_H.eps}
%\end{center}
%\caption{Switching current as a function of magnetic field for Al SNS dc squid 17ah.} 
%\label{17AlahIs_H} 
%\end{figure}

%\begin{figure}
%\begin{center} 
%\includegraphics[clip=true,width=7 cm]{17Alch_Is_H.eps}
%\end{center}
%\caption{Switching current as a function of magnetic field for Al SNS dc squid 17ch.} 
%\label{17AlahIs_H} 
%\end{figure}

\subsection{Conclusion}

The SNS dc squid resembles its SIS counterpart, with a strong modulation of its switching current by the magnetic field, with a periodicity of one flux quantum in the loop. But it also has unique features, such as a large switching current due to the low impedance of the normal metal. The IV curves and switching current histograms, because they depend on the intrinsic response functions of the normal metal wire, suggest the importance of investigating the dynamics of the Andreev states. In contrast to the SIS dc squid, the length and width of the normal part of the junctions can be varied at will. In the samples presented in this paper, the aspect ratio of the normal wires places them in the 1D or narrow 2D limit. We have shown that in this limit the decay of switching current at high field does not show an oscillating interference pattern typical of wider junctions, but instead decays monotonously in a Gaussianlike manner. Finally, some samples have a surprising increase of switching current at low field, which is still not well understood.

\subsection{Acknowledgements}
We acknowledge A. Anthore, M. Aprili, H. Courtois, P. Hekkila, F. Lefloch, I. Petkovic, H. Pothier, Ryazanov, B. Spivak for discussions, F. Pierre, D. Mailly and the LPN laboratory for help with the Nb samples fabrication in their clean room facilities.

\subsection{Appendix: Appearance of higher harmonics in a dc SNSquid}
The current phase relation in a SNS junction is not simply a sinusoidal relation but is predicted \cite{Heikkila} to contain higher harmonics:

\begin{equation}
I_{SNS}(\phi) = \sum_{n=1}^\infty I_c^n \sin(n\phi),
\label{current-phase}
\end{equation}
 with in the case of a long junction
\begin{equation}
I_c^n = -\frac{(-1)^n 33 e E_{Th}}{R_N(2n+1)(2n-1)}.
\end{equation}

The supercurrent through an SNS dc squid with two identical SNS junctions in parallel thus verifies:
\begin{eqnarray}
I_T=(\sum_{n} I_c^n\sin n(\delta_0 +\phi))+\sum_{n} I_c^n \sin n(\delta_0 -\phi)))\\
=2\sum_{n} I_c^n \sin n\delta_0 \cos n\phi),
\end{eqnarray}

where $\phi=e\Phi/\hbar$ and $\delta_0 +\phi$ and $\delta_0 -\phi$ are the phase differences at each of the junction.
Around $\delta_0=(2p+1)\frac{\pi}{2}$ (which is no longer the position of the maximum, but is close to it), it is easy to see that only the odd harmonics are non zero, so that the critical current is given by
\begin{equation}
I_c^{SNS}=2\sum_{n=2q+1}I_c^n \left|\cos n\phi \right|.
\end{equation}


\begin{thebibliography}{99}
\bibitem{temperature} Other experiments conducted on mesoscopic samples with the same apparatus same dilution refrigerator, same filtering), clearly saw temperature dependences down to 30 or 50 mK. However none of those samples consisted in normal wires connected to small superconducting contacts, for which heat dissipation is small).
\bibitem{Rqp} The resistance which should enter here is the quasiparticle resistance at frequency $\omega_p$. At this frequency which is of the order of the minigap induced in the metal, the quasiparticle energy is sufficient to overcome the minigap, and thus $R_{qp}\approx R_N$.
\end{thebibliography}
\end{document}